\documentclass[preprint,showpacs,preprintnumbers]{revtex4}
\usepackage{amssymb}
\usepackage{amsmath}
\usepackage{graphicx}
\usepackage{dcolumn}
\usepackage{bm}

\input{tcilatex}
\begin{document}

\title{Tunneling states in graphene heterostructures consisting of two
different graphene superlattices }
\author{Li-Gang Wang$^{1,2},$ Yuen-Chi Tse$^{2}$, and Shi-Yao Zhu$^{1,3}$}
\affiliation{$^{1}$Department of Physics, Zhejiang University, Hangzhou, 310027, China\\
$^{2}$Centre of Optical Sciences and Department of Physics, The Chinese
University of Hong Kong, Shatin, N. T., Hong Kong, China\\
$^{3}$Department of Physics, Hong Kong Baptist University, Kowloon Tong,
Hong Kong}

\begin{abstract}
We have theoretically investigated the properties of electronic transport in
graphene heterostructures, which are consisted of two different graphene
superlattices with one-dimensional periodic potentials. It is found that
such heterostructures possess an unusual tunneling state occurring inside
the original forbidden gaps, and the electronic conductance is greatly
enhanced and Fano factor is strongly suppressed near the energy of the
tunneling state. Finally we present the matching condition of the impedance
of the pseudospin wave for occuring the tunneling state by using the
Bloch-wave expansion method.
\end{abstract}

\email{wangligang@zju.edu.cn}
\maketitle

\section{Introduction}

Graphene has attracted a lot of attention since its discovery in 2004 \cite%
{Novoselov2004}. The linear dispersion property near the Dirac point leads
the low-energy charge carriers obeying the massless relativistic Dirac
equation. Thus many unique electronic and transport properties such as
half-integer quantum Hall effect \cite{Gusynin2005,Purewal2006} and Klein
paradox \cite{Katsnelson2006} are demonstrated in graphene.

Currently, most investigations are focused on the properties of graphene
itself. There are only a few works on graphene-based heterostructures, for
example, gated graphene heterostructures \cite{Ryzhii2006}, graphene
nanoribbon heterostructures \cite{Rosales2008}, and graphene/hexagonal-BN
heterostructures \cite{Bjelkevig2010}. Heterostructures are usually referred
to the combination of two different materials/structures \cite{Kroemer1963}.
According to the analog between the one-dimensional photonic crystals
containing left-handed metamaterials \cite{Li2003,JiangHT2003,Shadrivov2003}
and the graphene-based one-dimensional periodic superlattices \cite%
{Barbier2010,Arovas2010,Wang2010a}, many optical properties may be
introduced into the graphene-based electronic systems. In this paper, we
will theoretically study the graphene heterostructures consisted of two
different graphene-based superlattices (GSs), which can be constructed by
electrostatic \cite{Bai2007,Barbier2008,Park2008,Brey2009,Park2009,Park2008b}
or magnetic barriers \cite{RamezaniMasir2008,Ghosh2009}. Since the
electronic pseudospin wave may be highly localized at the interface between
two different GSs, it would be expected that such heterostructures may
possess an unusual tunneling mode inside the zero-averaged wave-number gap 
\cite{Wang2010a}. The location of such a tunneling state satisfies the
impedance matching condition of the electronic pseudospin wave.

\section{Theoretical Formula}

Now let us consider an electron of energy $E$ and wavevector $k_{0}=E/\hbar
v_{F}$ (with the Fermi velocity $v_{F}\approx 10^{6}$m/s) be incident at an
angle $\theta _{0}$ onto a graphene heterostructure $(AB)^{n}(CD)^{m}$, as
shown in Fig. 1, where $A$, $B$, $C$ and $D$ indicate different potentials,
and $n$ and $m$ are the period numbers. As we know that the Hamiltonian of a
low-energy electron in graphene is given by $\hat{H}_{j}=v_{F}\hat{\sigma}%
\cdot \hat{p}+V_{j}(x)\hat{I}$, here $\hat{p}=(-i\hbar \frac{\partial }{%
\partial x},-i\hbar \frac{\partial }{\partial y})$ is a momentum operator, $%
\hat{\sigma}=(\sigma _{x},\sigma _{y})$, $\sigma _{x}$ and $\sigma _{y}$ are
Pauli matrices, $\hat{I}$ is a $2\times 2$ unit matrix, and $V_{j}(x)$ is
the $j$th electrostatic potential. This Hamiltonian acts on a two-component
pseudospin state $\Psi =(\psi _{1},\psi _{2})^{T},$ where $\psi _{1}$ and $%
\psi _{2}$ are two components of the pseudospin wave (here we omit the term $%
e^{ik_{y}y}$ due to the translation invariance in the $y$ direction).
Therefore, the wavefunctions $\psi _{1,2}$ at $x$ and $x+\Delta x$ inside
the $j$th potential can be related via a transfer matrix: \cite{Wang2010a}%
\begin{equation}
M_{j}(\Delta x,E,k_{y})=\left( 
\begin{array}{cc}
\frac{\cos (k_{x,j}\Delta x-\theta _{j})}{\cos \theta _{j}} & i\frac{\sin
(k_{x,j}\Delta x)}{\cos \theta _{j}} \\ 
i\frac{\sin (k_{x,j}\Delta x)}{\cos \theta _{j}} & \frac{\cos (k_{x,j}\Delta
x+\theta _{j})}{\cos \theta _{j}}%
\end{array}%
\right) ,
\end{equation}%
where $k_{x,j}=$sign$(k_{j})\sqrt{k_{j}^{2}-k_{y}^{2}}$ is the $x$ component
of the wavevector $k_{j}=[E-V_{j}(x)]/\hbar v_{F}$ for $k_{j}^{2}>k_{y}^{2}$%
, otherwise $k_{x,j}=i\sqrt{k_{y}^{2}-k_{j}^{2}}$. Using the boundary
conditions, both the reflection and transmission coefficients could be
obtained by the transfer matrix method, \cite{Wang2010a}%
\begin{eqnarray}
r(E,k_{y}) &=&\frac{(x_{22}s_{0}-x_{11}s_{e})-x_{12}s_{0}s_{e}+x_{21}}{%
(x_{22}/s_{0}+x_{11}s_{e})-x_{12}s_{e}/s_{0}-x_{21}},  \label{rrcoeff} \\
t(E,k_{y}) &=&\frac{2\cos \theta _{0}}{%
(x_{22}/s_{0}+x_{11}s_{e})-x_{12}s_{e}/s_{0}-x_{21}},  \label{ttcoeff}
\end{eqnarray}%
where $s_{0}=e^{i\theta _{0}}$, $s_{e}=e^{i\theta _{e}}$, $\theta _{e}$ is
an exit angle (see Fig. 1), and $x_{ij}$ are the elements of the total
transfer matrix $\mathbf{X}\mathbf{=}\left( 
\begin{array}{cc}
x_{11} & x_{12} \\ 
x_{21} & x_{22}%
\end{array}%
\right) =\dprod\limits_{j=1}^{2n+2m}M_{j}(w_{j},E,k_{y})$ from the incident
to exit ends. Once the coefficients $r$ and $t$ are obtained, the total
conductance $G$ of the system at zero temperature can also be calculated by
using the B\"{u}ttiker formula \cite{Datta1995}, $G=G_{0}\int_{0}^{\pi
/2}T(E,k_{y})\cos \theta _{0}d\theta _{0}$, where $T=\left\vert
t(E,k_{y})\right\vert ^{2}$ is the transmittivity and $%
G_{0}=2e^{2}mv_{F}L_{y}/\hbar ^{2}$, and $L_{y}$ is the width of the
graphene strip in the $y$ direction. Furthermore, the Fano factor in this
system is given by \cite{Tworzydlo2006} $F=\left. \left( \int_{-\pi /2}^{\pi
/2}T(1-T)\cos \theta _{0}d\theta _{0}\right) \right/ \left( \int_{-\pi
/2}^{\pi /2}T\cos \theta _{0}d\theta _{0}\right) $. Therefore the
reflection, transmission, conductance, and Fano factor inside such
graphene-based heterostructures can be obtained by numerical calculations.

\section{Numerical Results}

Based on the above formulae, we can calculate the changes of the
transmission, conductance and Fano factor of an electron passing through the
heterostructures. In Fig. 2 and Fig. 3, we show the typical electronic
transmission, conductance and Fano factor for different situations with the
parameters $n=9$, $m=16$, $V_{A}=10$meV, $V_{B}=80$meV, $V_{C}=110$meV, $%
V_{D}=-20$meV, and $w_{A}=w_{B}=w_{C}=w_{D}=20$nm. For a single GS $(AB)^{n} 
$ or $(CD)^{m}$, from Fig. 2(a) and Fig. 3, the forbidden gaps do exist near
the new Dirac point at $E=45$meV, which is called the zero-averaged
wave-number gap in Ref. \cite{Wang2010a}, and the corresponding conductance
is minimal and the Fano factor equals to 1/3 at the new Dirac point of $E=45$%
meV for both cases of $(AB)^{n}$ and $(CD)^{m}$. Here we would like to point
out that the condition for appearing the Dirac point inside the GSs are
discussed in detail in Refs. \cite{Barbier2010,Arovas2010,Wang2010a}. For a
special case with $w_{A}=w_{B}$ and $w_{C}=w_{D}$, the condition is simply
given by $E=(V_{A}+V_{B})/2$ and $E=(V_{C}+V_{D})/2,$ derived from Eq. (41)
in Ref. \cite{Wang2010a}.

However, for the heterostructure $(AB)^{n}(CD)^{m}$, see Fig. 2(b), a new
tunneling mode occurs inside the forbidden gap near the Dirac point. Due to
the existence of this tunneling mode, compared with the case of the
individual $(AB)^{n}$ or $(CD)^{m}$, for the heterostructure the electronic
conductance is greatly enhanced and the Fano factor is strongly suppressed,
as shown in Figs. 3(a) and 3(b). In fact, this new tunneling mode is due to
the formation of the highly localized state at the interface between $%
(AB)^{n}$ and $(CD)^{m}$, which could be clearly seen from the density
distributions of the electronic pseudospin wavefunctions in Fig. 4. The
probability densities are highly localized at the interface between $%
(AB)^{n} $ and $(CD)^{m}$, which is denoted by the vertical dot line.

\section{Matching condition of the tunneling mode in forbidden gaps}

In order to obtain the condition of this new tunneling mode, we use the
so-called matching condition of the impedance of the pseudospin wave. In
infinite periodic barriers, $\Psi $ obeys the Bloch-Floquet theorem. By
using the Bloch-wave expansion method in Ref. \cite{Kang2009}, inside the GS 
$(AB)^{n}$ or $(CD)^{m}$, $\Psi _{AB,CD}$ can be written as 
\begin{equation}
\Psi _{p}(x)=H_{+}^{p}\left( 
\begin{array}{c}
1 \\ 
Z_{+}^{p}%
\end{array}%
\right) e^{i\beta _{x}^{p}x}+H_{-}^{p}\left( 
\begin{array}{c}
1 \\ 
Z_{-}^{p}%
\end{array}%
\right) e^{-i\beta _{x}^{p}x},  \label{BlochWave1}
\end{equation}%
where $\beta _{x}^{p}$ is the $x$ component of the pseudospin Bloch-wave
vector and satisfies the eigen equation: $Q_{p}\Psi _{p}=e^{\pm i\beta
_{x}^{p}\Lambda _{p}}\Psi _{p}$ [$Q_{p}$ is the matrix of the unit cell for $%
(AB)^{n}$ or $(CD)^{m}$ with $p=AB$ or $CD$], and $\Lambda _{p}$ is the
length of unit cell. $\left( 1,Z_{\pm }^{p}\right) ^{T}$ are the
eigenvectors of $Q_{p}$ corresponding to the eigenvalues $e^{\pm i\beta
_{x}^{p}D_{p}}$, and $H_{\pm }^{p}$ are the amplitudes of the forward (+)
and backward (-) Bloch waves. From Ref. \cite{Wang2010a}, $\Psi $ at both
the incident and exit regions can be, respectively, expressed by $\Psi
(0^{-})=\Phi _{i}(1+r)\left( 1,\xi _{0^{-}}\right) ^{T},$ and $\Psi
(x_{e}^{+})=t\Phi _{i}\left( 1,\xi _{x_{e}^{+}}\right) ^{T}$, where $\xi
_{0^{-}}=(e^{i\theta _{0}}-re^{-i\theta _{0}})/(1+r)$ and $\xi
_{x_{e}^{+}}=e^{i\theta _{e}}$ are the impedances at the incident and exit
ends, and $\Phi _{i}$ is the incident wavefunction. Inside the GS $(AB)^{n}$%
, from Eq. (\ref{BlochWave1}), we can define $\Psi _{AB}(0^{+})\equiv
H_{0^{+}}\left( 1,\xi _{0^{+}}\right) ^{T}$ and $\Psi _{AB}(x_{D}^{-})\equiv
H_{x_{D}^{-}}\left( 1,\xi _{x_{D}^{-}}\right) ^{T},$where $\xi _{0^{+}}$ is
the impedance at $x=0^{+}$ and $\xi _{x_{D}^{-}}$ is the impedance at the
interface $x=x_{D}^{-}$ ($x_{D}=n\Lambda _{AB}$), and $H_{0^{+}}$($%
H_{x_{D}^{-}}$) is a coefficient. Therefore, we can obtain 
\begin{equation}
\xi _{x_{D}^{-}}=\frac{Z_{+}^{AB}(\xi _{0^{+}}-Z_{-}^{AB})\lambda
_{AB}^{2}-Z_{-}^{AB}(\xi _{0^{+}}-Z_{+}^{AB})}{(\xi
_{0^{+}}-Z_{-}^{AB})\lambda _{AB}^{2}-(\xi _{0^{+}}-Z_{+}^{AB})},
\end{equation}%
with $\lambda _{AB}=e^{in\beta _{x}^{AB}\Lambda _{AB}}$. Similarly, the
impedance of the pseudospin wave at $x=x_{D}^{+}$ inside the GS $(CD)^{m}$
is given by%
\begin{equation}
\xi _{x_{D}^{+}}=\frac{Z_{+}^{CD}(\xi
_{x_{e}^{-}}-Z_{-}^{CD})-Z_{-}^{CD}(\xi _{x_{e}^{-}}-Z_{+}^{CD})\lambda
_{CD}^{2}}{(\xi _{x_{e}^{-}}-Z_{-}^{CD})-(\xi
_{x_{e}^{-}}-Z_{+}^{CD})\lambda _{CD}^{2}},
\end{equation}%
where $\lambda _{CD}=e^{im\beta _{x}^{CD}\Lambda _{CD}}$. For a perfect
tunneling mode in heterostructures, when and only when the impedances of the
pseudospin waves at two sides of the interface between $(AB)^{n}$ and $%
(CD)^{m}$ satisfy 
\begin{equation}
|\xi _{x_{D}^{-}}-\xi _{x_{D}^{+}}|=0,  \label{cond1}
\end{equation}%
the zero reflection can be exactly reached. In this case, the reflection
coefficient is also zero, i. e., $r=0$, thus $\xi _{0^{+}}=\xi
_{0^{-}}=e^{i\theta _{0}}$ at the incident end, and $\xi _{x_{e}^{-}}=\xi
_{x_{e}^{+}}=e^{i\theta _{e}}$ at the exit end. For the imperfect tunneling
process, the condition of Eq. (\ref{cond1}) should be replaced by $|\xi
_{x_{D}^{-}}-\xi _{x_{D}^{+}}|\rightarrow \min $. In Fig. 2(b), we plot the
curve for the value of $|\xi _{x_{D}^{-}}-\xi _{x_{D}^{+}}|$, see the dashed
line. From Fig. 2(b), it is clear that when the value of $|\xi
_{x_{D}^{-}}-\xi _{x_{D}^{+}}|$ is close to zero (within the original
forbidden gaps), there is a tunneling peak with near-unity transmittance.
According to the analog between one-dimensional photonic crystal
heterostructures \cite{Kang2009} and the graphene-based heterostructure
consisted of two different superlattices, we could safely say that in our
case the tunneling state is the electronic Tamm state. It should be
mentioned that, for the other zeroes of $|\xi _{x_{D}^{-}}-\xi _{x_{D}^{+}}|$
(which are located outside of the original forbidden gaps), they correspond
to the resonances of the propagation states, rather than the localized
states.

So far we only consider the GSs consisted of square barriers and wells. In
fact, the above results about the tunneling state will also occur inside the
graphene heterostructures, which are consisted of other periodic potentials
such as cosine or sine potentials \cite{Brey2009}. For a comparison, in Fig.
5, we also plot the electronic transmittance for the heterostructures
consisted of different sine potentials. It is clearly seen that the new
tunneling state also appears inside the original forbidden gaps under the
optimal period numbers.

\section{Conclusion}

In summary, we have investigated the electronic transport inside the
graphene-based heterostructures consisted of two different GSs. It is shown
that in such heterostructures there is an unusual tunneling state inside the
zero-averaged wave-number gaps, which leads to the enhancement of the
electronic conductance and the suppression of the Fano factor. The reason of
occurring such a tunneling state is due to the highly localization of the
electronic pseudospin wave at the interface between two GSs. Finally the
matching condition of the impedance of the pseudospin wave is analytically
presented. The parameters used in our study are technically feasible as
demonstrated in experimental studies \cite%
{Meyer2008,Marchini2007,VazquezdeParga2008,Pan2009}. For example, in Ref. 
\cite{Meyer2008}, superlattice patterns with periodicity as small as 5 nm
have been imprinted on graphene using the electron-beam induced deposition.
The phenomenon for occurring such a tunneling state in graphene
heterostructures is potentially useful in filtering single-energy electron
and is also hopefully to use in the design of graphene-based electronic
devices \cite{Peres2010}.

\begin{acknowledgments}
This work is supported by the National Natural Science Foundation of China
(No. 61078021) and Hong Kong RGC Grant No. 403609.
\end{acknowledgments}

\newpage

\begin{center}
{\LARGE Figure Captions}
\end{center}

Fig. 1. (Color online) (a) Schematic of a heterostructure consisted of two
different graphene superlattices, $(AB)^{n}$ and $(CD)^{m}$. (b) The
schematic profiles of the potentials.

Fig. 2.(Color online) Electronic transmittances through (a) the individual
graphene superlattice $(AB)^{n}$ or $(CD)^{m}$ and (b) the heterostructure $%
(AB)^{n}(CD)^{m}$, with $\theta _{0}=10^{\circ }$. In (b), the dashed line
denotes the value of $|\xi _{x_{D}^{-}}-\xi _{x_{D}^{+}}|$. The parameters
are $V_{A}=10$meV, $V_{B}=80$meV, $V_{C}=110$meV, $V_{D}=-20$meV, $%
w_{A}=w_{B}=w_{C}=w_{D}=20$nm, $n=9$, and $m=16$.

Fig. 3. (Color online) Changes of (a) conductance and (b) Fano factor in
different structures. Other parameters are the same as in Fig. 2.

Fig. 4.(Color online) Changes of probability densities $|\psi _{1,2}|^{2}$
inside the heterostructure, at the energy of the tunneling mode, $E\approx
44.688$ meV, and $\theta _{0}=10^{\circ }$. Other parameters are the same as
in Fig. 2.

Fig. 5. (Color online) Electronic transmittances through different
structures with the incident angle $\theta _{0}=10^{\circ }$: Case 1, $%
V_{AB}(x)=45-(35\pi /2)\sin (\pi x/20)$meV for $0<x<360$nm; Case 2, $%
V_{CD}(x)=V_{CD}(x)=45+(65\pi /2)\sin (\pi x/20)$meV for $0<x<1080$nm; Case
3, the combination of Cases 1 and 2.

\newpage

\begin{figure}[b]
\centering
\includegraphics[width=10cm]{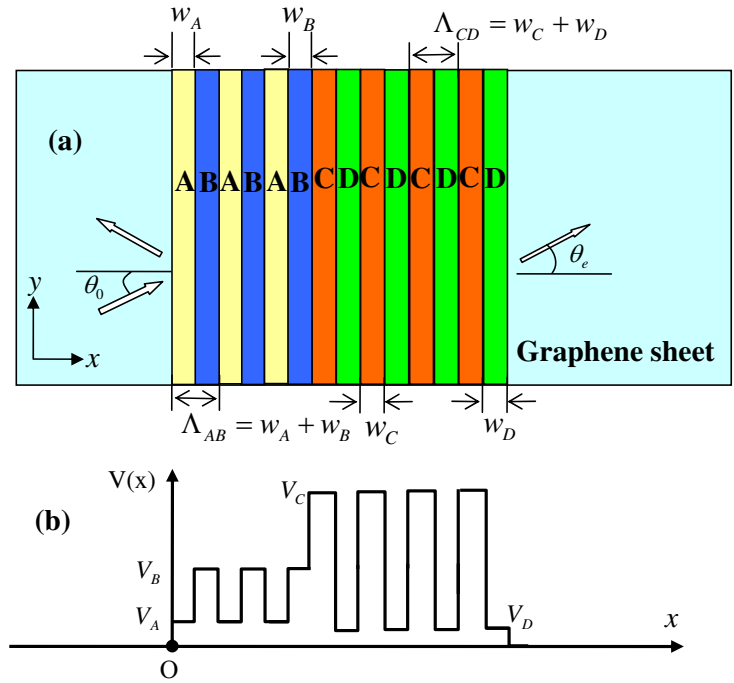}
\caption{(Color Online) (a) Schematic of a heterostructure consisted of two
different graphene superlattices, $(AB)^{n}$ and $(CD)^{m}$. (b) The
schematic profiles of the potentials.}
\label{fig:FIG1}
\end{figure}

\newpage

\begin{figure}[b]
\centering
\includegraphics[width=10cm]{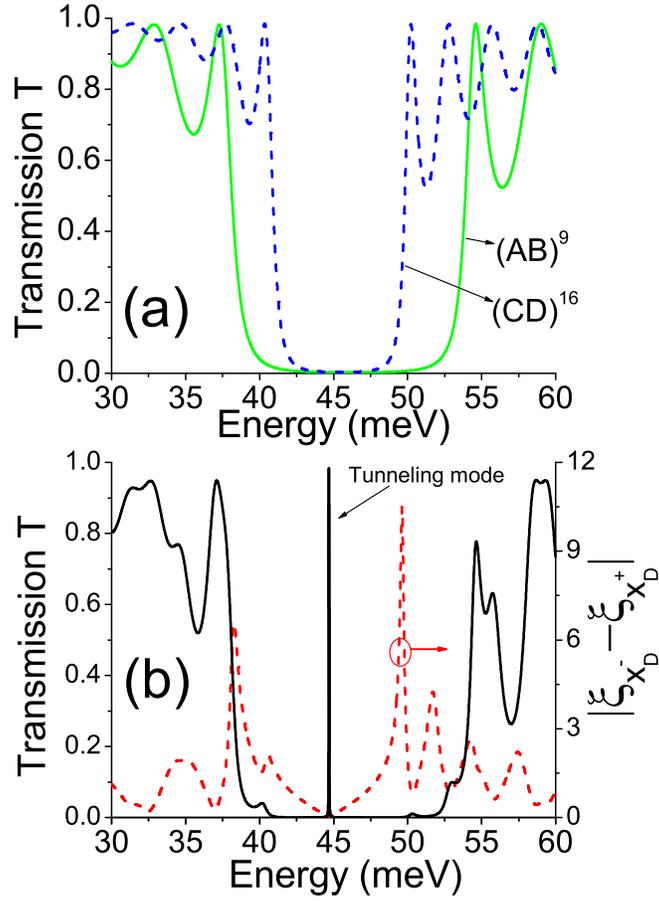}
\caption{(Color online) Electronic transmittances through (a) the individual
graphene superlattice $(AB)^{n}$ or $(CD)^{m}$ and (b) the heterostructure $%
(AB)^{n}(CD)^{m}$, with $\protect\theta _{0}=10^{\circ }$. In (b), the
dashed line denotes the value of $|\protect\xi _{x_{D}^{-}}-\protect\xi %
_{x_{D}^{+}}|$. The parameters are $V_{A}=10$meV, $V_{B}=80$meV, $V_{C}=110$%
meV, $V_{D}=-20$meV, $w_{A}=w_{B}=w_{C}=w_{D}=20$nm, $n=9$, and $m=16$.}
\label{fig:FIG2}
\end{figure}

\newpage

\begin{figure}[b]
\centering
\includegraphics[width=10cm]{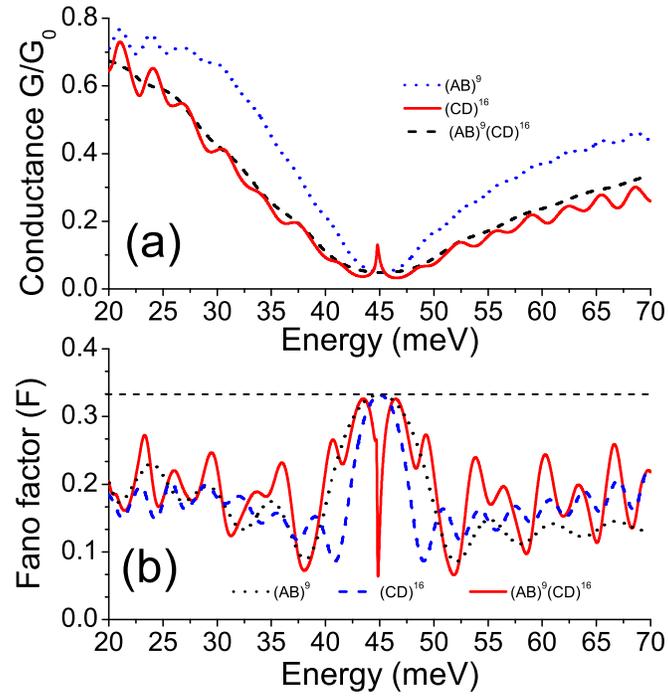}
\caption{(Color online) Color online) Changes of (a) conductance and (b)
Fano factor in different structures. The structral parameters are the same
as in Fig. 2. }
\label{fig:FIG3}
\end{figure}

\newpage

\begin{figure}[b]
\centering
\includegraphics[width=10cm]{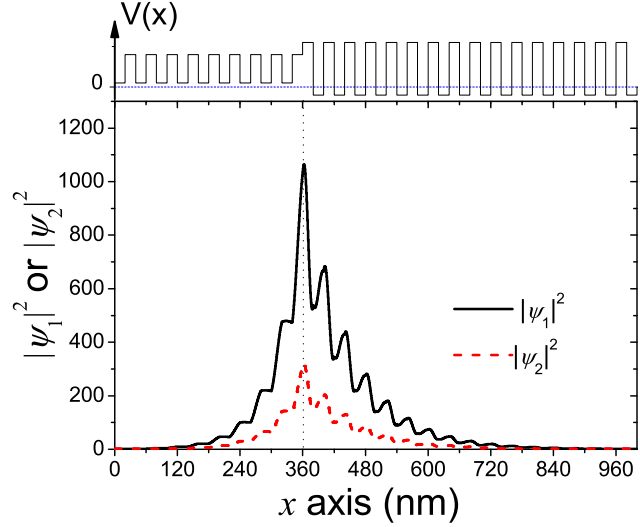}
\caption{(Color online) Changes of probability densities $|\protect\psi %
_{1,2}|^{2}$ of the pseudospin wave inside the graphene-based
heterostructure, at the energy of the tunneling mode, $E\approx 44.688$ meV
and $\protect\theta _{0}=10^{\circ }$. The vertical dot line denotes the
interface between $(AB)^{9}$ and $(CD)^{16}$. The other parameters are the
same as in Fig. 2.}
\label{fig:FIG4}
\end{figure}

\begin{figure}[b]
\centering
\includegraphics[width=10cm]{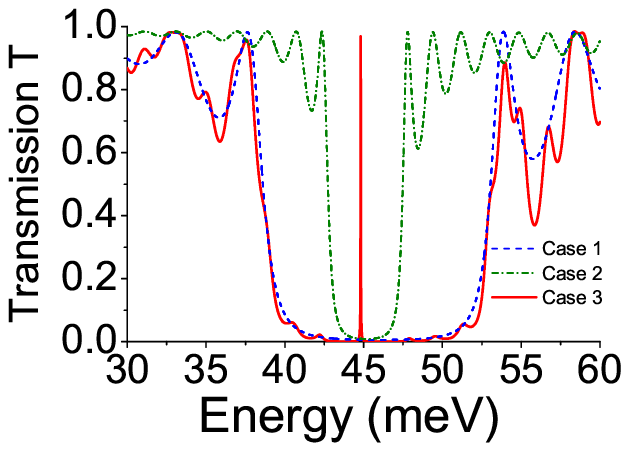}
\caption{(Color online) Electronic transmittances through different
structures with the incident angle $\protect\theta _{0}=10^{\circ }$: Case
1, $V_{AB}(x)=45-(35\protect\pi /2)\sin (\protect\pi x/20)$meV for $0<x<360$%
nm; Case 2, $V_{CD}(x)=V_{CD}(x)=45+(65\protect\pi /2)\sin (\protect\pi %
x/20) $meV for $0<x<1080$nm; Case 3, the combination of Cases 1 and 2.}
\label{fig:FIG5}
\end{figure}


\begin{thebibliography}{99}
\bibitem{Novoselov2004} K. S. Novoselov, A. K. Geim, S. V. Morozov, D.
Jiang, Y. Zhang, S. V. Dubonos, I. V. Grigorieva, and A. A. Firsov, Science, 
\textbf{306}, 666(2004).

\bibitem{Gusynin2005} V. P. Gusynin and S. G. Sharapov, Phys. Rev. B \textbf{%
71}, 125124 (2005).

\bibitem{Purewal2006} M. S. Purewal, Y. Zhang and P. Kim, Phys. Status
Solidi B \textbf{243}, 3418 (2006).

\bibitem{Katsnelson2006} M. I. Katsnelson, K. S. Novoselov and A. K. Geim,
Nature Physics \textbf{2}, 620 (2006).

\bibitem{Ryzhii2006} V. Ryzhii, Jpn. J. Appl. Phys. \textbf{45}, L923
(2006); V. Ryzhii, VA. Sato, and T. Otsuji, J. Appl. Phys. \textbf{101},
024509 (2007).

\bibitem{Rosales2008} L. Rosales, P. Orellana, Z. Barticevic, M. Pacheco,
MicroElectronics Journal, \textbf{39}, 537-540 (2008).

\bibitem{Bjelkevig2010} C. Bjelkevig, Z. Mi, J. Xiao, P. A. Dowben, L. Wang,
W. N. Mei, and J. A. Kelber, J. Phys.: Condens. Matter \textbf{22}, 302002
(2010).

\bibitem{Kroemer1963} H. Kroemer, Proceedings of the IEEE, vol. \textbf{51},
pp.1782-1783 (1963).

\bibitem{Li2003} J. Li, L. Zhou, C. T. Chan, and P. Sheng, Phys. Rev. Lett. 
\textbf{90}, 083901 (2003).

\bibitem{JiangHT2003} H. T. Jiang, H. Chen, H. Q. Li, Y. W. Zhang and S. Y.
Zhu, Appl. Phys. Lett. \textbf{83}, 5386 (2003).

\bibitem{Shadrivov2003} I. V. Shadrivov, A. A. Sukhorukov, and Y. S.
Kivshar, Appl. Phys. Lett. \textbf{82}, 3820 (2003).

\bibitem{Barbier2010} M. Barbier, P. Vasilopoulos, and F. M. Peeters, Phys.
Rev. B. \textbf{81}, 075438 (2010).

\bibitem{Wang2010a} L.-G. Wang and S.-Y. Zhu, Phys. Rev. B \textbf{81},
205444 (2010).

\bibitem{Arovas2010} D. P. Arovas, L. Brey, H. A. Fertig, E.-A. Kim, and K.
Ziegler, arXiv:1002.3655V2 (2010).

\bibitem{Bai2007} C. Bai and X. Zhang, Phys. Rev. B \textbf{76}, 075430
(2007).

\bibitem{Barbier2008} M. Barbier, F. M. Peeters, P. Vasilopoulos, and J.
Milton Pereira, Jr., Phys. Rev. B \textbf{77}, 115446 (2008).

\bibitem{Park2008} C. -H. Park, L. Yang, Y.-W. Son, M. L. Cohen and S. G.
Louie, Nature Physics \textbf{4}, 213 (2008).

\bibitem{Brey2009} L. Brey and H. A. Fertig, Phys. Rev. Lett. \textbf{103},
046809 (2009).

\bibitem{Park2009} C.-H. Park, Y.-W. Son, L. Yang, M. L. Cohen, and S. G.
Louie, Phys. Rev. Lett. \textbf{103}, 046808 (2009).

\bibitem{Park2008b} C.-H. Park, L. Yang, Y.-W. Son, M. L. Cohen, and S. G.
Louie, Phys. Rev. Lett. \textbf{101}, 126804 (2008)

\bibitem{RamezaniMasir2008} M. Ramezani Masir, P. Vasilopoulos, A. Matulis,
and F. M. Peeters, Phys. Rev. B \textbf{77}, 235443 (2008).

\bibitem{Ghosh2009} S. Ghosh and M. Sharma, J. Phys.: Condens. Matter 
\textbf{21,} 292204 (2009).

\bibitem{Datta1995} S. Datta, \textit{Electronic Transport in Mesoscopic
Systems}, Cambridge University Press, 1995.

\bibitem{Tworzydlo2006} J. Tworzyd\l o, B. Trauzettel, M. Titov, A. Rycerz,
and C. W. J. Beenakker, Phys. Rev. Lett.\textbf{\ 96}, 246802 (2006).

\bibitem{Kang2009} X. B, Kang, W. Tan, Z. G. Wang, and H. Chen, Phys. Rev. A 
\textbf{79}, 043832 (2009).

\bibitem{Meyer2008} J. C. Meyer, C. O. Girit, M. F. Crommie, and A. Zettl,
Appl. Phys. Lett. \textbf{92}, 123110 (2008).

\bibitem{Marchini2007} S. Marchini, S. G\"{u}nther, and J. Wintterlin, Phys.
Rev. B \textbf{76}, 075429 (2007).

\bibitem{VazquezdeParga2008} A. L. Vazquez de Parga, F. Calleja, B. Borca,
M. C. G. P. Jr, J. J. Hinarejo, F. Guinea, and R. Miranda, Phys. Rev. Lett. 
\textbf{100}, 056808 (2008).

\bibitem{Pan2009} Y. Pan, H. Zhang, D. Shi, J. Sun, S. Du, F. Liu, and H.-J.
Gao, Adv. Mater. \textbf{21}, 2777 (2009).

\bibitem{Peres2010} N. M. R. Peres, Rev. Mod. Phys. \textbf{82}, 2673 (2010).
\end{thebibliography}
\end{document}